# Minimal TestCase Generation for Object-Oriented Software with State Charts

Ranjita Kumari Swain[1], Prafulla Kumar Behera[2] and Durga Prasad Mohapatra[3]

[1]Rourkela Institute of Mgt. Studies, Rourkela
ranjita762001@yahoo.com
[2]Dept. of Comp. Sc., Utkal University, Bhubaneswar
P_behera@hotmail.com
[3]Dept. of Comp. Sc. and Engg., National Institute of Technology, Rourkela
durga@nitrkl.ac.in

*ABSTRACT*

*Today statecharts are a de facto standard in industry for modeling system behavior. Test data generation is one of the key issues in software testing. This paper proposes an reduction approach to test data generation for the state-based software testing. In this paper, first state transition graph is derived from state chart diagram. Then, all the required information are extracted from the state chart diagram. Then, test cases are generated. Lastly, a set of test cases are minimized by calculating the node coverage for each test case. It is also determined that which test cases are covered by other test cases. The advantage of our test generation technique is that it optimizes test coverage by minimizing time and cost. The present test data generation scheme generates test cases which satisfy transition path coverage criteria, path coverage criteria and action coverage criteria. A case study on Railway Ticket Vending Machine (RTVM) has been presented to illustrate our approach.*

## 1. INTRODUCTION

With continually increasing system sizes, the issue of automatic design of system test cases is assuming prime importance [28]. A properly generated test suite may not only locate the errors in a software system, but also help in reducing the high cost associated with software testing [17]. Many present day software solutions are state based. In such systems, the system behavior is determined by its state. In other words, a system can respond differently to the same event in different states. Therefore, unless a system is made to assume all its possible states and tested, it would not be possible to uncover state-based bugs. Adequate system testing of such software requires satisfactory coverage of system states and transitions. Generation of test specifications to meet these coverage criteria can be accomplished by using the state model of a system. However, it is a non-trivial task to manually construct the state model of a system. The state model of an actual system is usually extremely complex and comprises of a large number of states and transitions. Possibly for this reason, state models of complete systems are rarely constructed by system developers [28].

Testing of object-oriented software is a traditional activity that is part of the process of software quality assurance. Model-based testing technique has grown its importance. The models used represent the relevant features of the system under tests (SUT), and can also be used as a basis for generating test cases. The time and required effort to do sufficient testing grow, as the size and complexity of the software grows. Typical models that are used for representing system behavior are for instance the unified modeling language, finite state machines and state charts. Testing can be carried out earlier in the development process so that the developer will be able to find the inconsistencies and ambiguities in the specification. Hence it will be able to improve the specification before the program is written [12]. Testing activities consist of designing test cases that are sequences of inputs, executing the program with test cases, and examining the results produced by this execution. Unified modeling language (UML) has emerged as the de facto standard for modeling software systems and has received significant attention from researchers as well as practitioners.

UML models are popular not only for designing and documenting systems, the importance of UML models in designing test cases has also been well recognized. Even though UML models are intended to help reduce the complexity of a problem, with the increase in product sizes and complexities, the UML models themselves become large and complex involving thousands of interactions across hundreds of objects [16]. The important part of quality control in the software life-cycle is testing. As the complexity and size of software increase, the time and effort required to do sufficient testing grow. Manual testing is time-consuming and error-prone. So, there is a pressing to automate the testing process. The testing process can be divided into three parts: test case generation, test execution, and test evaluation. The latter two parts are relatively easy to automate provided that the criteria for passing the tests are available. However, to determine which tests are required to achieve a certain level of confidence is not trivial [18]. Model-based testing [6] has grown in importance. Models are specified to represent the relevant, desirable features of the system under consideration (SUC). These models are used as a basis for (automatically) generating test cases to be applied to the SUC. Typical models that are used for representing system behavior are unified modeling language, finite state machines, statecharts etc.[4]. It is often desired that test data in the form of test sequences within a test suite can be automatically generated to achieve required test coverage. It helps mankind to save on time, money and helps in handling better situations in real time.

With this motivation, we aim our work at deriving the test sequence from state transition diagram and maximizing state or node coverage. Also our method minimizes the size of test, time and cost, while preserving test coverage. The rest of the paper is structured as follows: A brief discussion on UML diagrams, which are relevant to our paper is described in the Section 2. Then, we discuss some testing coverage criteria in Section 3. Section 4 represents some concepts, notations and definitions of state chart diagram. In Section 5, we explain the overview of our proposed method for construction of state-transition graph, generation of test sequence using state charts and how node coverage is calculated for each test case. Section 6 provides the working of our methodology with the RTVM (Railway Ticket Vending Machine) case study and implementation of our example. Section 7 discusses some comparison with related work. Finally, Section 8 presents the conclusion and future work of this paper.

## 2. AN OVERVIEW OF RELEVANT UML 2.0 DIAGRAMS

In this section, we discuss an overview of the UML diagrams, which will be used subsequently in our paper.
UML is a modeling language using text and graphical notation and used for documenting specification, analysis, design, and implementation. It is a de- facto standard in industrial
software development UML, Unified Modelling Language is a visual language that has been developed to support the design of complex object-oriented systems. Since its introduction in
the late 90s, it has undergone several revisions. The latest release being UML version 2.0, which adds several new capabilities to UML 1.x. In this section, we restrict our review to only those developments that are directly relevant to our work. A typical software procedure incorporates all the three aspects: It uses data structure (class model), it sequences operations in time (state model), and it passes data and control among objects (interaction model). The three kinds of models separate a system into different views. The different models are not completely independent but a system is more than a collection of these independent parts. UML specification defines two major kinds of UML diagram: *structural diagrams* and *behavioral diagrams*. The elements in a structural diagram represent the meaningful concepts of a system, and may include abstract, real world and implementation concepts. Behavioral diagrams show the dynamic behavior of the objects in a system, which can be described as a series of changes to the system over time.

## 2.1. UML state chart diagram

In this section, we explain the few fundamentals on state chart diagram. The name of the diagram itself clarifies the purpose of the diagram and other details. It describes different
states of a component in a system. The states are specific to a component or object of a system. A statechart diagram describes a state machine. Now to clarify it state machine can be defined as a machine which defines different states of an object and these states are controlled by external or internal events. Statechart diagram is one of the five UML diagrams used to model dynamic nature of a system. They define different states of an object during its lifetime. And
these states are changed by events. So, Statechart diagrams are useful to model reactive systems. Reactive systems can be defined as systems that respond to external or internal events. Statechart diagram describes the flow of control from one state to another state. States are defined as a condition in which an object exists and it changes when some event is triggered. So, the most important purpose of Statechart diagram is to model life time of an object from creation to termination. Statechart diagrams are also used for forward and reverse engineering of a system. But the main purpose is to model reactive system.

State diagrams are used to give an abstract description of the behavior of a system. This behavior is analyzed and represented in series of events, that could occur in one or more possible states. Hereby "each diagram usually represents objects of a single class and track the different states of its objects through the system". State diagrams can be used to graphically represent finite state machines. Followings are the main purposes of using statechart diagrams: (a) To model dynamic aspect of a system, (b) To model life time of a reactive system and (c) To describe different states of an object during its life time.

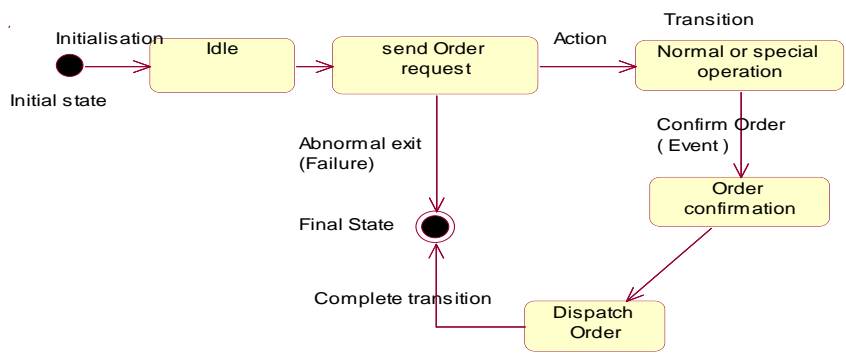

Figure. 1. Simple state chart for an order object in Order Management System

The fundamental components of a state chart diagram are - events, states and transitions. But it has some distinguishing characteristics for modeling dynamic nature. State chart diagram defines the states of a component and these state changes are dynamic in nature. So, its specific purpose is to represent the state changes triggered by events. Events are internal or external factors influencing the system. If we look into the practical implementation of Statechart diagram then it is mainly used to analyze the object states influenced by events.
This analysis is helpful to understand the system behavior during its execution. The following are the basic notational elements that can be used to make up a diagram:

**2.1.1. State:** The state of an object is shown by rectangle with rounded corners. Top of the rectangle contains a name of the state. It can contain a horizontal line in the middle, below which the activities that are done in that state are indicated. A state in a state chart diagram can either be simple or composite type. A simple state does not have any sub-states.

**2.1.2. Initial state:** A transition leading from an initial event shows the state that an object goes into when it is created or initialized. This is shown as a small black disk or filled circle. A state-chart can have only one initial state.

**2.1.3. Final state:** Like initial state the state diagram shows final state. It represents the state reached when an object is destroyed, switched off or stops responding to events. This is shown as a hollow circle containing a smaller filled circle or small black disk within a large circle. A state-chart may have more than one final state [24].

**2.1.4. Transition:** Arrows, denote transitions. The name of the event (if any) causing this transition is written as the labels with the transition names or event names. A guard expression may be added before a "/" and enclosed in square-brackets ( *eventName[guardExpression]* ), denoting that this expression must be true for the transition to take place.

**2.1.5. An event:** We call the event that causes the state transition the trigger. An event is an occurrence at a point of time. Events often correspond to verbs in the past tense e.g. (power turned on, alarm set) [5]. There are four types of events that can trigger a state transition:

*Signal event (when the system receives a signal from an external agent)*

*Call event (when a system operation is invoked)*

*Timing event (when a timeout occurs)*

*Change event (when a system property is changed by an external agent)*

**2.1.6. Choice and Junction Points:** A choice point allows a transition to branch to several different states depending on the value of a guard. A junction point indicates that several states can transition to the same state on a given event.

## 3. SOME BASIC DEFINITIONS

Here, in this section, we introduce a few basic concepts and definitions which we use while describing our aproach in the subsequent sections. There are many forms of state diagrams,
which differ slightly and have different semantics. For a deterministic finite state machine (DFA), nondeterministic finite state machine (NFA), generalized nondeterministic finite state machine (GNFA), Mealy machine or Moore machine, the input is denoted on each edge. For a Mealy machine, input and output are signified on each edge, separated with a slash "/":
"1/0" denotes the state change upon encountering the symbol "1" causing the symbol "0" to be output. For a Moore machine the state's output is usually written inside the state's circle, also separated from the state's designator with a slash "/". There are also variants that combine these two notations.
It is assumed that a state chart STc is correct in the sense that for each state $s \in S_{simple}$ there exists a sequence of transitions $t_1, t_2..., t_k$ so that source($t_1$) $\in Si$ and target($t_k$) = s and for each state s ∈ $S_{simple}$ there exists a sequence of transitions $t_1$, $t_2$..., $t_k$ so that source($t_1$) = s and target($t_k$) ∈ $S_f$. The following terms will be used to describe our technique.

***Definition 1.*** A statechart can be a quadruple $S_c$ = (E, $S_t$, H, T), where E is a finite set of events and $St = (S, S_i, S_f)$ is a triple of set of states with S as a finite set of states, $S_i ⊆ S$ denoting the entries (initial states) and $S_f ⊆ S$ the exits (final states),

H ⊆ S × S is a hierarchy relation, a binary relation on the set S forming a tree. For an element (s, s′) ∈ H holds, that a state s is an immediate sub state of state S′.

T ⊆ S × E × S is a finite set T of transitions. The set of states S is composed of disjoint sets of simple states $S_{simple}$ and composite states $S_{comp}$.

*Scenario-intermediate state:* A scenario is executed, when the system is at a state (*Si*) may cause the system to transit from its current state (*Si*) to a next state (*Sj*). Since a scenario may consist of a number of message exchanges among objects, a system may even change its state during the execution of a step (a message) in a scenario. Hence, execution of a single scenario may cause traversal of several states and transitions. All system states other than *Si* and *Sj*, through which the system transits during the execution of a scenario are considered to be scenario-intermediate state. We can say, execution of a scenario may cause a system to transit from one state to another state, and all other intermediate states reached during execution of the scenario are scenario-intermediate states. It may be important that a scenario-intermediate state reached during execution of a scenario can also be a normal state, if it is either the initial or final state for execution of some scenario.

***Definition 2.*** A transition pair TP = (t, t′) with t, t′ belongs to $T_{legal}$ is a sequence of a legal incoming transition to a legal outgoing transition of a (simple) state so that ∃ s ∈ $S_{simple}$:

t ∈ in(s) ∪ t′ ∈ out(s).

*Transitions and scenario-intermediate transitions:* Let S = $St_1$, $St_2$, . . . , $St_n$ be the set of all possible system states of an SUT. We define a transition *tij* to be a tuple ($St_i$, S*, $St_j$), where $t_{ij}$ represents a transition from the current state $St_i$ ∈ S to a next state $S_{tj}$ ∈ S, S* is an ordered set of all he scenario-intermediate states reached during the transition from state $St_i$ to $St_j$. A transition for which either the source or destination state is a scenario-intermediate state, is known as a scenario-intermediate transition.

***Definition 3.*** A false transition pair FP = (t, t′) with t ∈ $T_{legal}$ and t′ ∈ $T_{faulty}$ is a sequence of a legal incoming transition to a faulty outgoing transition of a (simple) state so that ∃ s ∈ $S_{simple}$ :
t ∈ in(s) ∪ t′ ∈ out(s).

***Definition 4.*** A sequence of n legal transitions ($t_1$, $t_2$..., $t_n$) with $t_i$ ∈ $T_{legal}$ where ($t_i$, $t_{i+1}$) denotes a valid transition pair for all i ∈ 1, ..., n - 1 is called a transition sequence ($T_{seq}$) of length n. A transition sequence ($t_1$, $t_2$..., $t_n$) is complete if it starts at the initial state of the state chart that is entered firstly and ends at a final state. In this case it is called a complete transition sequence ($T_{sqcom}$).

***Definition 5.*** A fault transition sequence $T_{sqfault}$ = ($t_1$, $t_2$, ..., $t_n$) of length n consists of n - 1 subsequent transitions, forming a (legal) transition sequence of the length n - 1 plus a concluding, faulty transition $t_n$ ∈ $T_{faulty}$. A faulty transition sequence is called complete if it starts at the initial state of the statechart, abbreviated as CFTS. The sequence ($t_1$, $t_2$, ..., $t_{n-1}$) is called a start sequence.

***Definition 6.*** A test case is the triplet [I, S, O], where I is the initial state of the system at which the test data is input, S is the test data input to the system and O is the expected output of the system [18], [19]. The output produced by the execution of the software with a particular test case provides a specification of the actual software behavior. A test case is also characterized by an ordered pair of an input and an expected output of the SUC.

*Definition 7*. A test suite is a set of test cases. A single test case in most cases may satisfy more than one test obligation. For instance, a test case used to cover a certain state of interest may also cover other states during its execution. This then provides for a way to reduce the size of the final test suite by choosing a subset of test cases that preserves the coverage obtained by the full test suite [9].

*Definition 8.* k-transition coverage (k-TC) generates complete transition sequences that sequentially conduct all legal transition sequences of length k $\in$ N.

*Definition 9.* A single test case in most cases may satisfy more than one test obligation. For instance, a test case used to cover a certain state of interest may also cover other states
during its execution. This then provides for a way to reduce the size of the final test suite by choosing a subset of test cases that preserves the coverage obtained by the full test suite [10].
A test suite is a set of test cases.

Definition 8 guarantees that all possible (legal) transition sequences of length k will be tested. A test suite consisting of all transition sequences of a fixed length k does not necessarily cover a set of all sequences of length i $\in$ 1, ... , k-1 as there may exist sequences of length i that cannot be expanded to length k. Typically, state based test generation methods focus on some form of coverage, for instance on covering transitions [19], [20] or on transition coverage and state identification [6], [14]. Our approach creates all transition sequences of length k including all shorter sequences of length 1, ..., k -1 that cannot be found in longer sequences.

## 4. TEST ADEQUECY CRITERIA

Typically, state based test generation methods focus on some form of coverage, for instance on covering transitions [20], [21] or on transition coverage and state identification [7], [15]. Our approach creates all transition sequences of length k including all shorter sequences of length 1, ..., k -1 that cannot be found in longer sequences. Testing coverage/adequacy criterion specifies the requirement of a particular testing and can be used as an objective measurement of the test case. In traditional software code testing, the definition of testing adequacy is defined as a measurement function. The case of UML state chart diagrams is different because it is in the form of model instead of code. Especially the coverage of state chart diagram is little bit complex. In our paper, we propose different types of coverage metrics as follows:

- *State Coverage Criterion :* The value of state coverage is the ratio between the covered states and all the states in the statechart diagram. It requires that all the state nodes in a state chart graph to be covered at least once.
- *Action Coverage*: To generate test cases with respect to action coverage we construct as many marked specifications as there are actions within the specification.

- *Transition Coverage :* Given a graph and a test suite TS, *s* is said to achieve transition coverage, if it causes each transition *t* of the state graph to be exercised at least once. The value of transition coverage is the ratio between the checked transitions and all the transitions in the statechart graph.

- *Transition Path Coverage:* Given a graph and a test suite *TS*, s is said to achieve transition path coverage, if it causes each elementary transition path *p* of the state graph to be exercised at least once. The value of path coverage is the ratio between the traversed paths and all the paths in the graph.

- *Condition coverage.* A single condition is covered, if it evaluates to both true and false at some point during test execution. Decision coverage has also been called branch coverage or predicate coverage. This means that 100% condition/decision coverage is achieved if all conditions evaluate to true and to false and if every decision also evaluates to true and to false during the test execution. A decision consists of conditions separated by logical operators (e.g. and, or).

## 5. TEGEMIOOSC–OUR PROPOSED APPROACH TO GENERATE AND MINIMIZE TEST SEQUENCE WITH STATECHARTS

In this section we discuss the overview of our proposed approach to generate test sequence from UML state chart diagram and then, we optimize the test node coverage while minimizing time and cost. We have named our approach, *Test Generation and Minimization for O-O software with State Charts* (TeGeMiOOSc). The schematic representation of our approach is shown in Figure. 2. Our proposed methodology involves the following steps.

- Step-1. Analyze the real system which is to be tested and accepted by user.
- Step-2. Construct State Chart Diagram
- Step-3. Convert the given State Chart Diagram into an intermediate graph. We named this intermediate graph a state transition graph.
- Step-4. Starting from the first node, traverse remaining nodes, using DFS concept, in order to form test sequence.
- Step-5. Obtain all the valid sequences of the application until final edge is reached.
- Step-6. Minimize a set of test cases by calculating node coverage for each test sequence.

In the next Section, we discuss each step in detail, by using different algorithms for each step in different sections. Before that, we present some related definitions. In the next section, we discuss each step in detail, by using different algorithms for each step in different sections. Before that we present some related definitions.

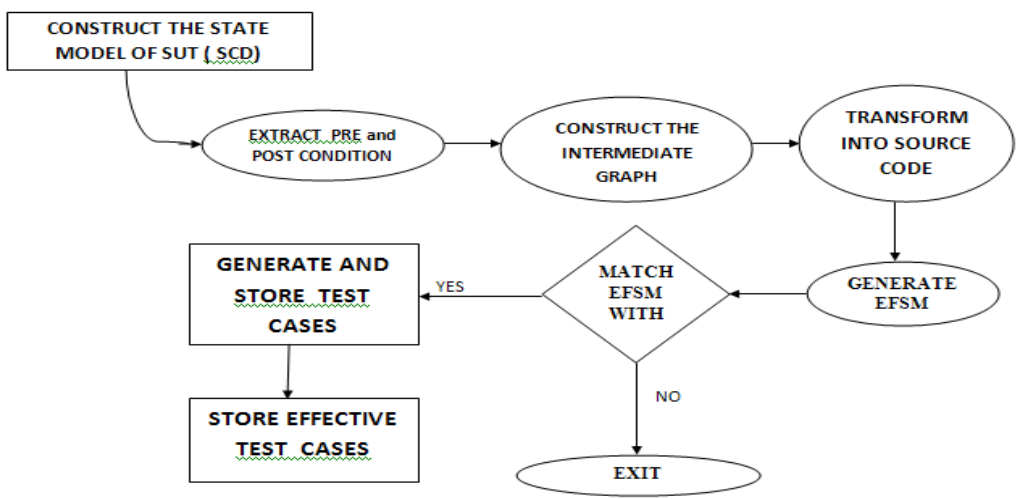

Fig. 2. Schematic representation of our approach

## 5.1. Analyze the real system which is to be tested

A Railway Ticket Vending System (RTVM) dispenses tickets to passengers at a railway station. passengers use the front panel to specify their Boarding and destination place, details of passengers(number of adults and children) and date of travel. The machine displays the fare for the requested ticket. The passengers deposits cash in the bin provided and presses 'accept cash'. The machine checks the cash, if it is more; the balance cash is paid out. And the ticket requested is printed. And the report options also include the detailed report of transactions, summary report of the number of tickets sold for each destination, opening balance, each collected, cash dispensed and current balance in the machine. The total functionalities are shown by the use case diagram as shown in Fig. 3.

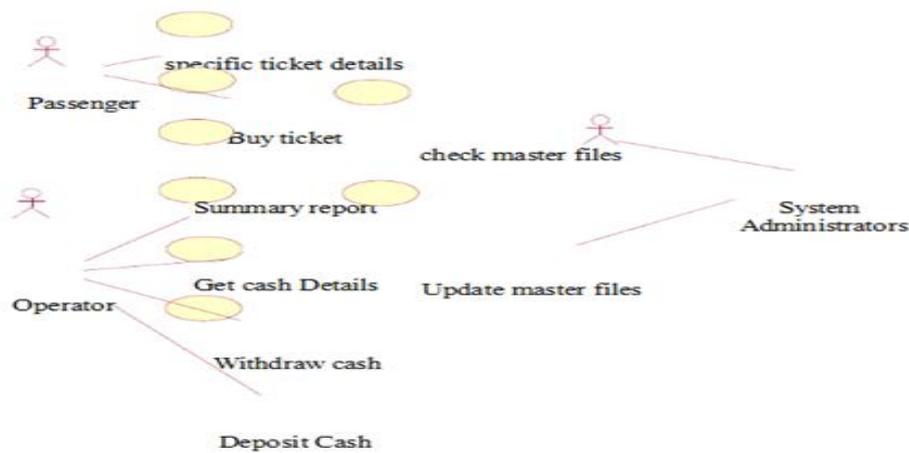

Figure. 3. Use case diagram for SUT

## 5.2. Construct the model of the SUT using state chart diagram
State chart diagram is a graph where nodes represent states end the directed arcs that interconnect states represent transitions. It also models dynamic behavior, and captures the different states that an object can be in, and its response to various events that may arise in each of its states. Statechart diagram is one of the 13 UML diagrams used to model dynamic nature of a system. They define different states of an object during its lifetime. The notation and semantics of UML state diagrams are substantially based on Statecharts modified to include object-oriented features [13]. The states and the transition of a system are important to set up a state diagrams from system. Fig. 4 shows a UML state diagram for a railway ticket vending machine system.

## 5.3. Convert the state chart diagram into state transition graph

Here, we convert the state diagram into state transition graph. A state transition graph $TG = (V_t, E_d)$.

***Definition 10:*** A **transition graph** $TG = (V_t, E_d)$ represents a directed graph consisting of a set of vertices ($V_t$), a set of directed edges ($E_d$). In TG, nodes represent states and edges represent transitions between states. Without any loss of generality, we assume that there is a unique node that corresponds to the initial state and that one or more nodes represent the final states. The initial state is represented as the root of the tree. States at each level of nesting are considered as a sub graph. We represent this step through an algorithm.

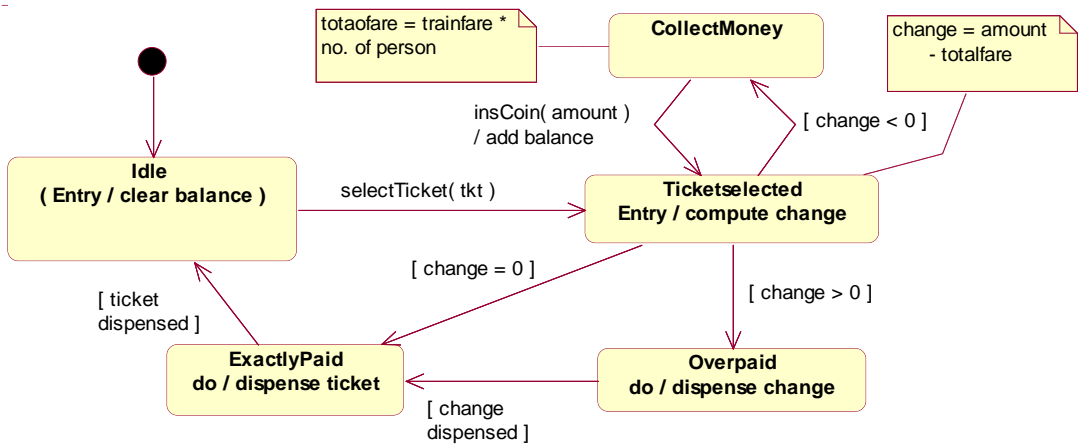

Figure. 3. State chart diagram for Railway Ticket Vending Machine

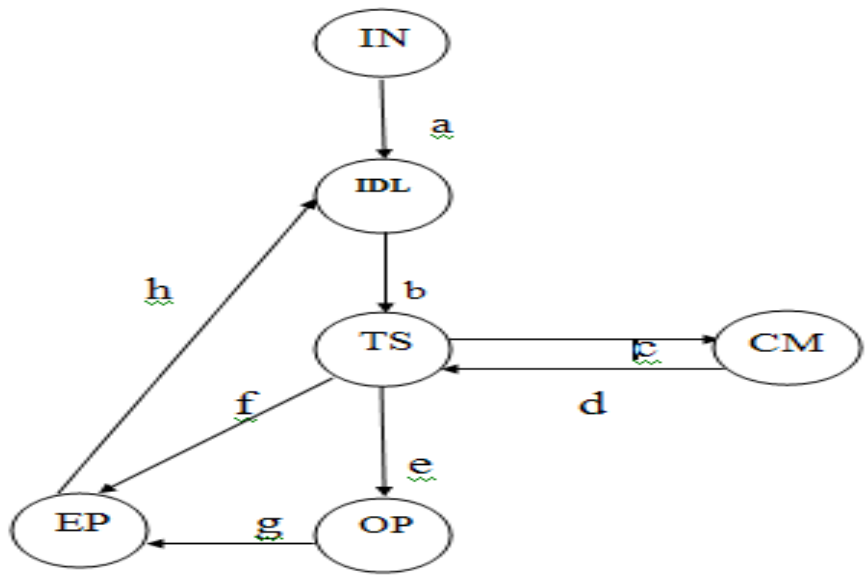

Fig. 5. Corresponding transition graph of the statechart diagram of RTVM

### 5.4. Traverse the state transition graph

Here, we present how the graph is traversed to extract all the information which are required to generate the test sequences. Here,
ST = {St1, St2, St3 ....Stn}, each Sti is a stage or node
ID = {ID1, ID2, ID3, . . . , IDn}, each IDi is an input data
OD = {OD1, OD2, OD3, . . . , ODn}, each ODi is an output value
TR = {TR1, TR2, TR3, . . . , TRn}, each TRi is a transition between source and destination stage, where each TRi = {Stp, Stq}, Stp is a source stage and Stq is a destination stage.

Set V = {t| t ∈ Tlegal} consists of vertices representing the legal transitions of statechart Sc. For each (legal) transition pair (t, t′) of the statechart, a directed edge is created. Vertex *ti* has to be

connected with all transitions that may be triggered from the state belonging to the initial configuration. Transitions leading into a final state have to be connected with vertex $tj$. A simple transition coverage may be reached by visiting all vertices of the graph, based on a transition graph at least once by starting in vertex ti and ending in vertex $tj$. The problem of computing a route for visiting all vertices of a graph by minimizing the length of the route is well known as the traveling salesman problem (TSP). If visiting vertices and traversing edges more than once is allowed, it is called the graphical traveling salesman problem (GTSP) [25]. Also, by computing all complete transition sequences whose length is smaller than k and that cannot be expanded to longer sequences, a minimal test case set fulfilling k-transition coverage for all k $\in$ 1, . . . , n is achieved. This procedure is described in Algorithm-1, in the next section.

### 5.4. Generating test cases

After, a transition graph is constructed based on a state chart *Sc*. If n = l, the graphical traveling salesman problem can be applied directly. If n is greater than 1 the transition graph has to be transformed k - 1 times. The resulting graph represents all possible sequences of transitions of length k. Additionally, all sequences of length k-1 are computed that cannot be expanded to longer sequences. These sequences are characterized by the fact that the corresponding vertex representing that sequence is solely connected with vertices $t_i$ and $t_j$. The functions indeg(v) := v′,(v′, v) $\in$ A and outdeg(v) := v′(v, v′) $\in$ A are used to compute these vertices. As these sequences are already complete, they can be added to the set TS.

**Algorithm - 1: Generation of a test case set for k-transition coverage for k $\in$ 1, ..., n**

**Input: A state transition graph from** *Sc* **= (E,*St*,H, T), n $\in$ N**
**Output: A test case set TS fulfilling k-transition coverage**

1. **For** k $\in$ 1, ..., n
2. **Begin**
3.     i:= 0; j:= 1; TS = null; s = null; visited NSt = 0
4.     **Do while** NSt [i] = NULL
5.     push ( NSt [i], s);
6.     visited NSt [ NSt [i] ] = visited NSt [ NSt [i] ] + 1;
7.      **Do while** s *not* = NULL
8.         t = pop(s); NSt[i] = pop(s);
9.         **If** enabled ( NSt[i]) not = NULL
10.           visitnexttransition( NSt [i] );
11.         **Endif**
12.     **Enddo**
13.     TS = $\phi$
14.     **For k** := 2 to n do
15.         **Foreach** v $\in$ V do
16.           **If** ($t_i$, v) $\in$ *Ed* $\wedge$ (v, $t_j$) $\in$ *Ed* $\wedge$ indeg(v) = outdeg(v) = 1 then
17.             TS := TS U v
18.             Ed := Ed U ($t_i$, $t_j$)
19.           Add them to set TS
20.           **Endif**
21.         **Endfor**
22.         **Enddo**
23.     **Enddo**
24.     **End**
25. **Endfor**

### 5.6. Minimizing the test cases

Here, in this section, we discuss how the generated test cases are reduced while maximizing test coverage. Though Wang's algorithm [17] is widely used, but it does not cover other critical attributes, like defect id, dependency and automated test case indicator. In order to generate an effective size of generated test cases, this step contains two sub activities, which are

- calculate node coverage for each test case. Let $NC(tc) = t_1, t_2,...t_n$ for $NC(tc)$ to be a set of test cases that tc is covered by $t_1, t_2,...t_n$. Hence, if a number of testset tc is zero, then tc is included in the effective set of test cases.

- select effective test cases. Now, we present this step through an algorithm.

    Now, we select the effective testcases, which contains no testset. We present this step through an algorithm.

### Algorithm-2 (*Minimization of Test cases*)

**Input:** $Sc = (E, St, H, T)$, Transition graph and A set of test cases TS[n];
**Output:** Reduced test suite RTs;
1. ANC, PNC: Node coverage, Previous coverage, tc: test case
2. RTs = ϕ;
3. $NC(tc) = t_1, t_2,...t_n$
4. ANC = 0;
5. repeat
6. select a test case F from TS[n];
7. **Foreach** $s \in S_{simple}$ do
8.     **Foreach** $t \in in(s)$ do
9.         **Foreach** $t' \in out(s) \cap T_{faulty}$ do
10.         ANC := ANC U (t, t′)
11.         **Foreach** $(t, t') \in ANC$ do
12.             $(t_1, ..., t_i)$ := STARTSEQUENCE(t)
13.             RTs := RTs U $(t_1, ...t_i, t, t')$
14.             **If** in(inital(root())) = ϕ then
15.                 RTs := RTs U (t)
16.                 select effective test cases;
17.                 **If** a number of set tc is zero;
18.                     **then** tc is included in the effective set of test cases;
19.                 Endif
20.             Endif
21.         Endfor
22.         Endfor
23.     Endfor
24. **Endfor**

## 6. CASE STUDY

### 6.1. Problem statement

Buying ticket in RTVM (Railway Ticket Information Systems) is a system which receives the cash from the customer and returns back the balance and the ticket. The system has five components: Front panel, Ticket Transacter, Cash Transacter, Cash Handler and Ticket. Entire functions of the system are represented in the form of a usecase diagram. The customer deposits the cash. The system then passes the cash deposit intimation to ticket transacter. The ticket transacter then accepts cash and passes it to cash transacter and the cash transacter then transacts the cash to the cash handler. After receiving the cash, the cash handler sends an acknowledgment to the cash transacter. The cash transacter gives cash confirmation acknowledgement to the ticket transacter and also gives the balance amount to cash transacter. The cash transacter sends the acknowledgment for the cash received to the ticket transacter. Finally the ticket transacter prints the ticket with the help of ticket component and issues the ticket to the customer. The state diagram for the buy ticket in RTVM system is shown in Fig. 4.

Here, the object enters into *idle* state, when the power switch is on. Once the user selects a tickettype button in the menu, the object enters the *ticketselected* state. The user can select the destination, ticket type and the number of persons (n) to travel. The condition n ≤ 6 is inserted for the event tickettypeselected, as the ticket machine is not expected to issue a ticket for more than 6 persons in one transaction. Once the ticket type and number of persons required are selected, the object enters the *collectmoney* state. In this state, the object collects the amount of money (totalfair) the user has to enter into the ticket machine. Note that totalfair = ticket fare × number of persons. As the user inserts money (amount) into the machine, the machine object changes its state to busy. In the busy state, it calculates how much balance or change (chng) has to be returned to the user if any, where chng = (amount - totalfair). If the change balance is less than zero, the machine object changes its state again from *busy* to *collectmoney* as the money inserted is insufficient. If the change balance is equal to zero, the machine object goes to *exactlypaid* state and dispenses the ticket for requested number of persons. If the change balance is more than zero, then the machine object changes its state to *overpaid* state and then dispenses ticket as well as dispenses change. After constructing state chart diagram, we construct transition graph as shown in Fig 5.

In next step, we traverse the graph and extract all required information from the state transition graph as described below.

### 6.2. Working Of the proposed Algorithm

The state model application saved as .mdl file is provided as input to the parser. The parser analyses and collects all the information about object states, actions / guards and transitions, which are represented as nodes and edges in a directed graph as shown in Fig. 5. In Table I, we represent the mapping of object states to its corresponding nodes in the graph and Table II shows actions or guard conditions to corresponding edges in the graph. Then, by using traversal algorithm we find all possible test sequences generated and are shown in Table III.

TABLE I
MAPPING INFORMATION TABLE FOR OBJECT STATES

| Object States in state model | Nodes in graph |
|---|---|
| Initial state | IN |
| Idle state | IDL |
| Ticketselected | TS |
| Collectmoney | CM |
| Overpaid | OP |
| Exactlypaid | EP |

Hence, the set of states are ST = {IN, IDL, TS, CM, OP, EP}, each $ST_i$ is a state or node
ID = {$ID_1$, $ID_2$, $ID_3$, $ID_4$, $ID_5$, $ID_6$}, each $ID_i$ is an input data OD = {$OD_1$, $OD_2$, $OD_3$, $OD_4$, $OD_5$, $OD_6$}, each $OD_i$ is an output value
TR = {$TR_1$, $TR_2$, $TR_3$, $TR_4$, $TR_5$, $TR_6$, $TR_7$}, each $TR_i$ is a transition between source and destination state, where each $TR_i$ = {$ST_p$, $ST_q$}, STp is a source state and $ST_q$ is a destination state. Hence, each transition can be extracted as follows:

TR1 = { IN, IDL }
TR2 = { IDL, TS }
TR3 = { TS, CM }
TR4 = { CM, TS }
TR5 = { TS, OP }
TR6 = { TS, EP }
TR7 = { OP, EP }
TR8 = { EP, IDL }

This step is to verify the completion of extracted information, derived from the diagram.

TABLE II
MAPPING TABLE FOR TRANSITION / ACTION / GUARD

| Transition/Action/Guard | corresponding edge in graph |
|---|---|
| Initialization | a |
| select ticket | b |
| G1: change < 0 | c |
| insert( coinamount ) | d |
| G2: change > 0 | e |
| G3: change = 0 | f |
| change dispensed | g |
| ticket dispensed | h |

Next, we derive and generate test cases. Hence, all tests can be generated as follows:
The last step is to minimize the set of test cases by calculating node coverage for each test case and determine which test cases are covered by other test cases.
NC(tc1) = {tc9, tc10, tc11, . . . tc19}
NC(tc2) = {tc9, tc10, tc11, . . . tc19}
NC(tc3) = {tc10, ... tc14}
NC(tc4) = {tc11, ... tc14}
NC(tc5) = {tc12, ... tc17}
NC(tc6) = {tc18, tc19}
NC(tc7) = {tc13, tc14, tc16, tc17 }
NC(tc8) = { tc14, tc17, tc19}
NC(tc9) = {tc10, tc11, . . . tc19}
NC(tc10) = {tc11, tc12, tc13, tc14}
NC(tc11) = {tc12, tc13, tc14}
NC(tc12) = {tc13, tc14}
NC(tc13) = {tc14 }
NC(tc14) = { }
NC(tc15) = {tc16}
NC(tc16) = {tc17}
NC(tc17) = { }
NC(tc18) = {tc19}
NC(tc19) = { }

Therefore, the following test cases such as tc1, tc2, tc3, tc4, tc6, tc8, tc9, tc10, tc11, tc12, tc15, tc16 should be ignored. Hence, the remaining effective test sequence is TS = {tc14, tc17, tc19}. The node coverage of the above test sequence are shown below in Table IV.

TABLE III
TESTCASE GENERATED TABLE FOR RTVM

| Test Sequence | Result |
|---|---|
| $tc_1$ = IN-(a)-IDL | valid |
| $tc_2$ = IDL-(b)-TS | valid |
| $tc_3$ = TS-(c)-CM | valid |
| $tc_4$ = CM-(d)-TS | valid |
| $tc_5$ = TS-(e)-OP | valid |
| $tc_6$ = TS-(f)-EP | valid |
| $tc_7$ = OP-(g)-EP | valid |
| $tc_8$ = EP-(h)-IDL | valid |
| $tc_9$ = IN-(a)-IDL-(b)-TS | valid |
| $tc_{10}$ = IN-(a)-IDL-(b)-TS-(c)-CM | valid |
| $tc_{11}$ = IN-(a)-IDL-(b)-TS-(c)-CM-(d)-TS | valid |
| $tc_{12}$ = IN-(a)-IDL-(b)-TS-(c)-CM-(d)-TS-(e)-OP | valid |
| $tc_{13}$ = IN-(a)-IDL-(b)-TS-(c)-CM-(d)-TS-(e)-OP-(g)-EP | valid |
| $tc_{14}$ = IN-(a)-IDL-(b)-TS-(c)-CM-(d)-TS-(e)-OP-(g)-EP-(h)-IDL | valid |
| $tc_{15}$ = IN-(a)-IDL-(b)-TS-(e)-OP | valid |
| $tc_{16}$ = IN-(a)-IDL-(b)-TS-(e)-OP-(g)-EP | valid |
| $tc_{17}$ = IN-(a)-IDL-(b)-TS-(e)-OP-(g)-EP-(h)-IDL | valid |
| $tc_{18}$ = IN-(a)-IDL-(b)-TS-(f)-EP | valid |
| $tc_{19}$ = IN-(a)-IDL-(b)-TS-(f)-EP-(h)-IDL | valid |

TABLE IV
CALCULATED NODE COVERAGE TABLE FOR RTVM

| Node coverage ( NC ) | Resultant Test Sequence |
|---|---|
| NC($tc_1$) | {tc9, tc10, … tc19} |
| NC($tc_2$) | {tc9, tc10, … tc19} |
| NC($tc_3$) | {tc10, ... tc14} |
| NC($tc_4$) | {tc11, ... tc14} |
| NC($tc_5$) | {tc12, ... tc17} |
| NC($tc_6$) | {tc18, tc19} |
| NC($tc_7$) | {tc13, tc14, tc16, tc17 } |
| NC($tc_8$) | { tc14, tc17, tc19} |
| NC($tc_9$) | {tc10, tc11, … tc19} |
| NC($tc_{10}$) | {tc11, ... tc14} |
| NC($tc_{11}$) | {tc12, ... tc14} |
| NC($tc_{12}$) | {tc13, tc14} |
| NC($tc_{13}$) | {tc14 } |
| NC($tc_{14}$) | { } |
| NC($tc_{15}$) | {tc16} |
| NC($tc_{16}$) | {tc17} |
| NC($tc_{17}$) | { } |
| NC($tc_{18}$) | {tc19} |
| NC($tc_{19}$) | { } |

### 6.3. Implementation

In this section, we present some experimental results in order to verify the effectiveness of our approach. We have carried out a series of experiments. In our experiments we have considered different types of applications. All these systems are designed in UML 2.0 using Rational Rose Sostware. Here, we consider the example of Railway Ticket Vending Machine (RTVM). We can experiment with other applicatins namely Library Information System (LIS), Cell Phone System (CPS), Trading House Automation System (TAS) etc.

First, the usecase diagram of RTVM is modelled using RationalRose software. then considering only the Purchase Ticket or Buy ticket usecase, we model it with State chart diagram and saved with .mdl extension. Next, the state chart diagram is converted into state chart graph. Now, the statechart graph is traversed applying DFS technique, considering pre and post conditions. Then, the graph is transformed into the sourcecode in JAVA. Hence, the designs are also implemented using JAVA and NetBeans IDE version 7.1.2. The screenshot of our Java sourcecode is shown in Fig.6.

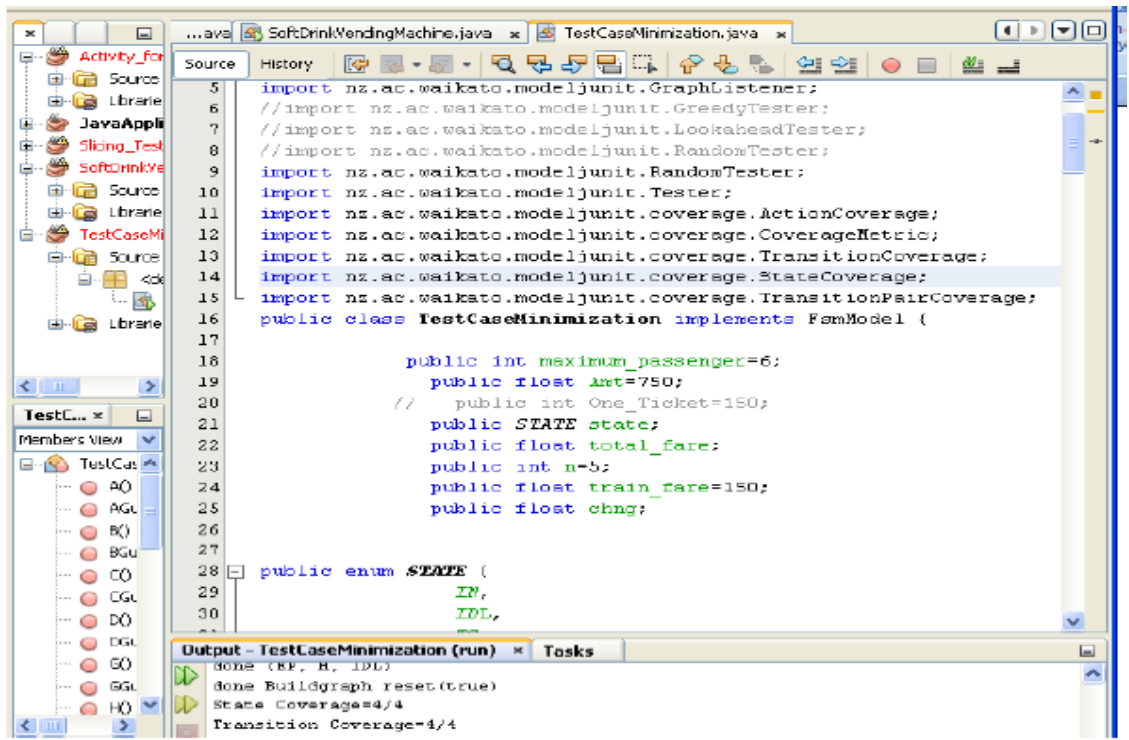

Fig. 6. Screenshot of JAVA source code of State chart graph

We have implemented our approach by using ModelJunit tool. ModelJUnit [1] is a Java library that extends JUnit to support model-based testing. It allows us to write simple FSM or EFSM models as Java classes, then generate tests from those models and measure various model coverage metrics. System models are extended finite state machines that are written in a familiar and expressive language: JAVA. ModelJUnit is an open source tool, which is released under the GNU GPL license. ModelJUnit is an openly available test case generation tool, using an extended finite state chart diagram. ModelJUnit searches the graph of an EFSM at runtime. It generates EFSM (Extended Finite State Machine) from source code. The screenshot of the generated EFSM for different conditions are shown in Fig. 7. EFSM models are graphically represented as graphs where states are represented as nodes and transitions as directed edges between states.

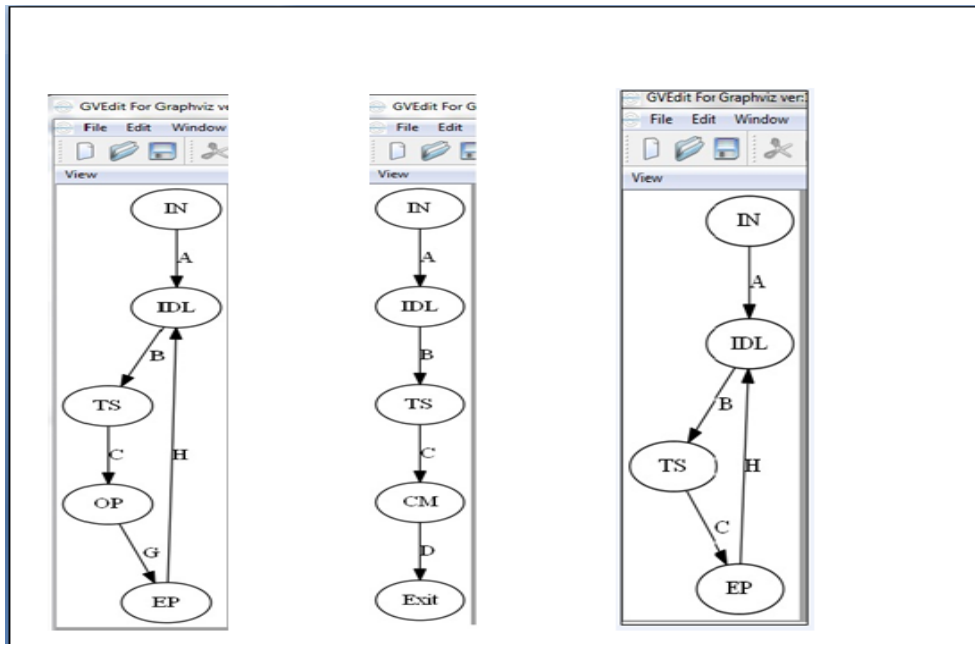

Figure. 7. EFSM generated for different conditions

The source and destination states as well as the prefix path conditions are displayed along with the test data. In our implementation, we have restricted the conditional expressions in state diagrams to have only integer and Boolean variables as these occur commonly. The GUI provides a friendly and efficient user interface to user to generate testing code and shows test coverages as shown in Fig. 8.

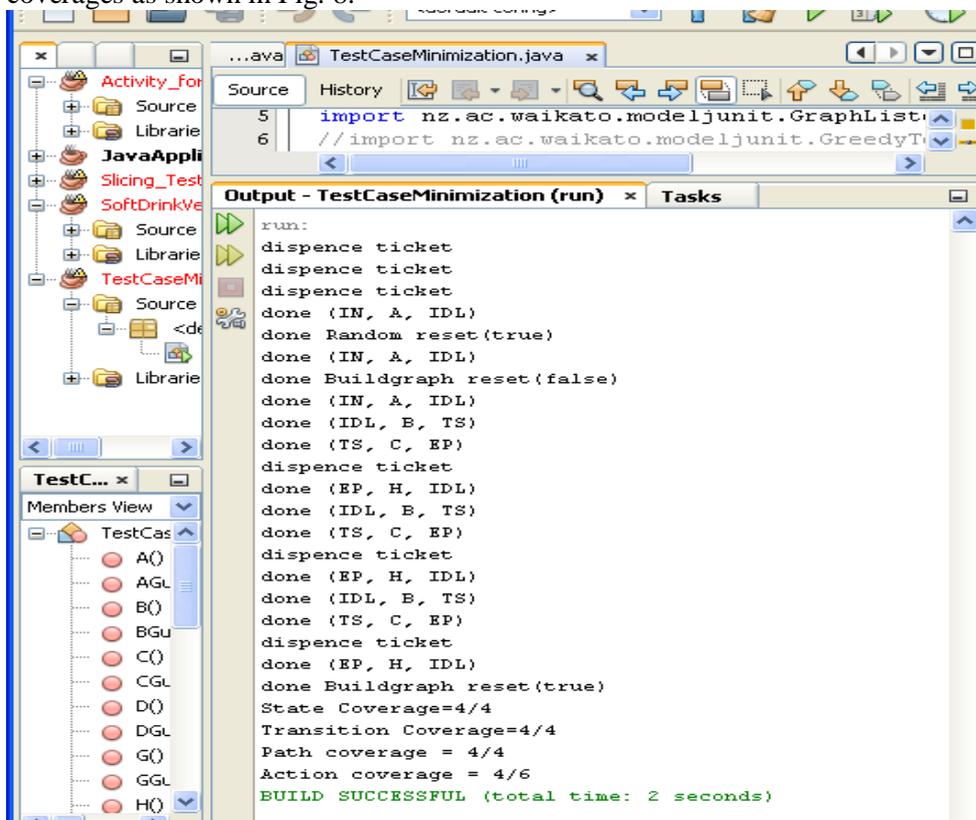

Figure 8. Test coverage through our implementation

We have implemented the sytem model by providing several inputs. Fig. 9 shows some coverage results for different values of input data. The used abbreviations have these following meanings:

NOT: No.Of Tickets, SC:STATE COVERAGE , TC: TRANSITION COVERAGE, PC: PATH COVERAGE, AC: ACTION COVERAGE

But, our achieved coverages are shown as follows

ASC: ACHIEVED STATE COVERAGE , ATC: ACHIEVED TRANSITION COVERAGE, APC: ACHIEVED PATH COVERAGE, AAC: ACHIEVED ACTION COVERAGE

| NOT | MONEY | CONDITION | SC | TC | PC | AC | ASC | ATC | APC | AAC |
|---|---|---|---|---|---|---|---|---|---|---|
| 2 | 200 | LESS MONEY | 5 | 4 | 3 | 6 | 5 | 4 | 3 | 4 |
| 3 | 500 | MORE MONEY | 5 | 5 | 5 | 6 | 5 | 5 | 5 | 5 |
| 4 | 500 | LESS MONEY | 5 | 4 | 3 | 6 | 5 | 4 | 3 | 4 |
| 5 | 700 | LESS MONEY | 5 | 4 | 3 | 6 | 5 | 4 | 3 | 4 |
| 6 | 1000 | MOREMONEY | 5 | 5 | 5 | 6 | 5 | 5 | 5 | 5 |
| 7 | 1000 | LESS MONEY | 5 | 4 | 3 | 6 | 5 | 4 | 3 | 4 |
| 4 | 600 | EXACT MONEY | 4 | 4 | 4 | 6 | 4 | 4 | 4 | 4 |
| 5 | 1000 | MOREMONEY | 5 | 5 | 5 | 6 | 5 | 5 | 5 | 5 |
| 7 | 500 | LESS MONEY | 5 | 4 | 3 | 6 | 5 | 4 | 3 | 4 |
| 5 | 1000 | MOREMONEY | 5 | 5 | 5 | 6 | 5 | 5 | 5 | 5 |

Figure . 9. Number of test coverage through different input values

### 6.4. Test Reduction

There is hardly sufficient time for thorough testing activities within industrial projects, in practical situations. Hence, it is reasonable to try to reduce the size of generated test suites. However, the effect of the reduction on the fault detection ability of the test suites should be small. The techniques proposed in our paper can be used to apply reduction during test case generation. A single test case may cover more than the coverage item it has been generated for. When using a probe based technique as described in our paper it is easy to identify all items covered by a particular test case.

# 7. COMPARISON WITH RELATED WORK

A lot of research work have been investigated the effect of testset reduction on the size and fault finding capability of a test-set. In an early study, Wong et al. address the question of the effect on fault detection of reducing the size of a test set while holding coverage constant [29], [30]. They randomly generated a large collection of test sets that achieved block and all-uses data flow coverage for each subject program. For each test set they created a minimal subset that preserved the coverage of the original set. They then compared the fault finding capability of the reduced test-set to that of the original set. Their data shows that test minimization keeping coverage constant results in little or no reduction in its fault detection effectiveness. This observation leads to the conclusion that test cases that do not contribute to additional coverage are likely to be ineffective in detecting additional faults.

To confirm or refute the results in the Wong study, Rothermel et al. performed a similar experiment using seven sets of C programs with manually seeded faults [26]. For their experiment they used edge-coverage [8] adequate test suites containing redundant tests and compared the fault finding of the reduced sets to the full test sets. In this experiment, they found that [3] the fault-finding capability was significantly compromised when the test-sets were reduced and [27] there was little correlation between test-set size and fault finding capability.
The results of the Rothermel study were also observed by Jones and Harrold in a similar experiment [11].

Prasanna M, et al.[23] proposed a technique for generating test cases using collaboration diagrams. They converted the diagram into an intermediate graph and from the graph, by applying Prim's and Dijkstra's algorithm, generated a set of test cases.

Offutt and Abdurazik [20], [21] proposes a technique for generating test cases from UML state diagrams. They have highlighted several useful test coverage criteria for UML
state charts such as: (1) full predicate coverage, (2) transition coverage etc.

Kansomkeat and Rivepiboon [12] introduce a method for generating test sequences using UML state chart diagrams. They transformed the state chart diagram into a flattened
structure of states called testing flow graph (TFG). From the TFG, they listed possible event sequences which they considered as test sequences. The testing criterion they used to guide the generation of test sequences is the coverage of the states and transitions of TFG.

Kim et al. [13] proposes a method for generating test cases for class testing using UML state chart diagrams. They transformed state charts to extended FSMs (EFSMs) to derive test cases. In the resulting EFSMs, the hierarchical and concurrent structure of states are flattened and broadcast communications are eliminated. Then data flow is identified by transforming
the EFSMs into flow graphs, to which conventional data flow analysis techniques are applied.

In the experiment discussed in our paper, we attempt to highlight some additional issues. These different results are difficult to reconcile and the relationship among coverage criteria, testsuite size, and fault finding capability clearly needs more study.

Our work is different in some respects. First, we are not studying testing of traditional programs. We generate testcases for object oriented software. we are interested in test case generation as well as testcase minimization by calculating node coverage for each test case and testing of specifications. A single test-case in most cases may satisfy more than one test obligation. For instance, a testcase used to cover a certain state of interest may also cover other states during its execution. It provides a way to reduce the size of the final test-suite by choosing a subset of test-cases that preserves the coverage obtained by the full testsuite.

# 8. CONCLUSSION AND FUTURE WORK

In our approach, first, we built a state chart model of our system under test. Next, we derived state transition graph from state chart diagram. Then, all required information are extracted from the graph. Next, the test cases are generated by our algorithm. Lastly, a set of test cases are minimized by calculating node coverage for each test case. It is determined
that which test cases are covered by other test cases. In this way, our paper introduces an efficient test generation approach to optimize the test coverage by minimizing time and cost. In our opinion, there is almost no degradation in terms of testsuite quality.

Further, we plan into testcase generation techniques by using any other UML diagram such as activity diagram. Experimental studies required to determine if such techniques
can more reliably reduce the burden of the testing effort.

**Authors**

**Ranjita Ku. Swain** completed her MCA from College of Engg.And Technology, OUAT, Bhubaneswar, India. She is pursuing her Ph.D degree from Utkal University, Vani vihar, Bhubaneswar, India. She is currently working as Senior Lecturer in Computer Science Dept., Rourkela Institute of Management Studies, Rourkela, India. She has 11years of teaching experience and her fields of interest are Software Engg., Discrete Mathematical Structure and Numerical Methods.

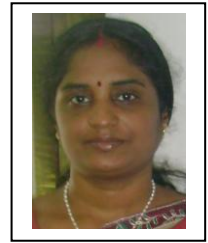

**Prafulla Ku. Behera** has received his Ph.D degree from Utkal University, Vani vihar, Bhubaneswar, India. He is currently working as a reader at Dept. of Computer Science & Application, in Utkal University, Vani vihar, Bhubaneswar, India. His special fields of interest include Mobile Computing, Software Engineering, He is a member of CSI.

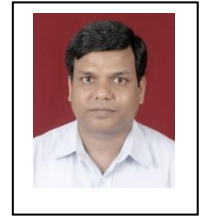

**Durga Prasad Mohapatra** received his Ph. D. from Indian Institute of Technology Kharagpur and M. E. from Regional Engineering College (now NIT), Rourkela. He joined the faculty of the Department of Computer Science and Engineering at the National Institute of Technology, Rourkela in 1996, where he is now Associate Professor. His research interests include software engineering, real-time systems, discrete mathematics and distributed computing and published more than forty papers in these fields. He has received many awards including Young Scientist Award for the year 2006 by Orissa Bigyan Academy, Prof. K. Arumugam award for innovative research for the year 2009 and Maharasthra State National Award for outstanding research for the year 2010 by ISTE, NewDelhi. He has also received three research projects from DST and UGC. Currently, he is a member of IEEE. Dr. Mohapatra has co-authored the book *Elements of Discrete Mathematics: A computer Oriented Approach* published by Tata Mc-Graw Hill.

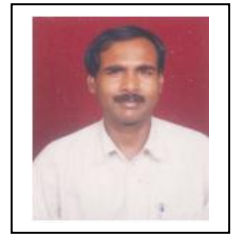